\documentclass[pra,final,floatfix,twocolumn]{revtex4-1}
\pdfoutput=1

\usepackage{graphicx}
\usepackage{dcolumn}
\usepackage{bm}
\usepackage{amssymb}
\usepackage{natbib}
\usepackage{amsmath}
\usepackage{color}
\usepackage{datetime}
\usepackage{footnote}
\usepackage{bbold}
\usepackage{braket}
\usepackage[normalem]{ulem}
\usepackage{dcolumn}
\usepackage{ragged2e}
\usepackage{xfrac}
\usepackage{float}
\usepackage{sidecap}

\newcommand{\gradi}[0]{^{\circ}}

\newcommand{\be}{\begin{equation}}
\newcommand{\ee}{\end{equation}}

\newcommand{\nocontentsline}[3]{}
\newcommand{\tocless}[2]{\bgroup\let\addcontentsline=\nocontentsline#1{#2}\egroup}

\newif\iffigs
\figstrue

\usepackage[breaklinks=true]{hyperref}

\hypersetup{
  colorlinks   = true, 
  urlcolor     = blue, 
  linkcolor    = blue, 
  citecolor   = red 
}

\begin{document}

\title{Experimental Demonstration of Self-Guided Quantum Tomography}

\author{Robert J. Chapman$^{1}$}

\author{Christopher Ferrie$^{2}$}

\author{Alberto Peruzzo$^{1}$}
\email{alberto.peruzzo@rmit.edu.au}

\affiliation{
$^{1}$Quantum Photonics Laboratory, School of Engineering, RMIT University, Melbourne, Australia and\\ School of Physics, The University of Sydney, Sydney, Australia\\
$^{2}$Centre for Engineered Quantum Systems, School of Physics, University of Sydney, Sydney, Australia\\
}

\begin{abstract}
Robust, accurate and efficient quantum tomography is key for future quantum technologies. Traditional methods are impractical for even medium sized systems and are not robust against noise and errors. Here we report on an experimental demonstration of self-guided quantum tomography; an autonomous, fast, robust and precise technique for measuring quantum states with significantly less computational resources than standard techniques. The quantum state is iteratively learned by treating tomography as a projection measurement optimization problem. We experimentally demonstrate robustness against both statistical noise and experimental errors on both single qubit and entangled two-qubit states. Our demonstration provides a method of full quantum state characterization in current and near-future experiments where standard techniques are unfeasible.
\end{abstract}

\maketitle

Quantum technologies require high fidelity preparation, control and characterization of quantum states, for application in quantum metrology \cite{giovannetti_advances_2011}, quantum simulators \cite{lloyd_universal_1996}, and quantum computers \cite{ladd_quantum_2010}. Recent advances in quantum control of several qubits have enabled demonstrations of quantum error correction \cite{nigg_quantum_2014} and boson sampling \cite{broome_photonic_2013,spring_boson_2013}. Standard quantum tomography (SQT) has been the cornerstone of quantum state characterization for decades \cite{stokes_composition_1851,fano_description_1957,banaszek_focus_2013}. As the number of quantum particles increases, the total number of measurements and computation resources required to characterize a quantum state and store its parameters grow exponentially. SQT has the additional computational cost of performing a state-estimation inverse problem and requires, for example, maximum likelihood estimation to avoid unphysical results. The scaling and additional computational cost make SQT impractical for quantum states being prepared today \cite{vlastakis_deterministically_2013,haas_entangled_2014}. The reliability of SQT for all system sizes is limited by sensitivity to statistical noise and experimental errors. Unless modified at additional resource cost \cite{branczyk_self-calibrating_2012}, SQT fails in the presence of measurement errors \cite{enk_when_2013}.

\begin{figure}[t]
\centering
\includegraphics[width=0.7\linewidth]{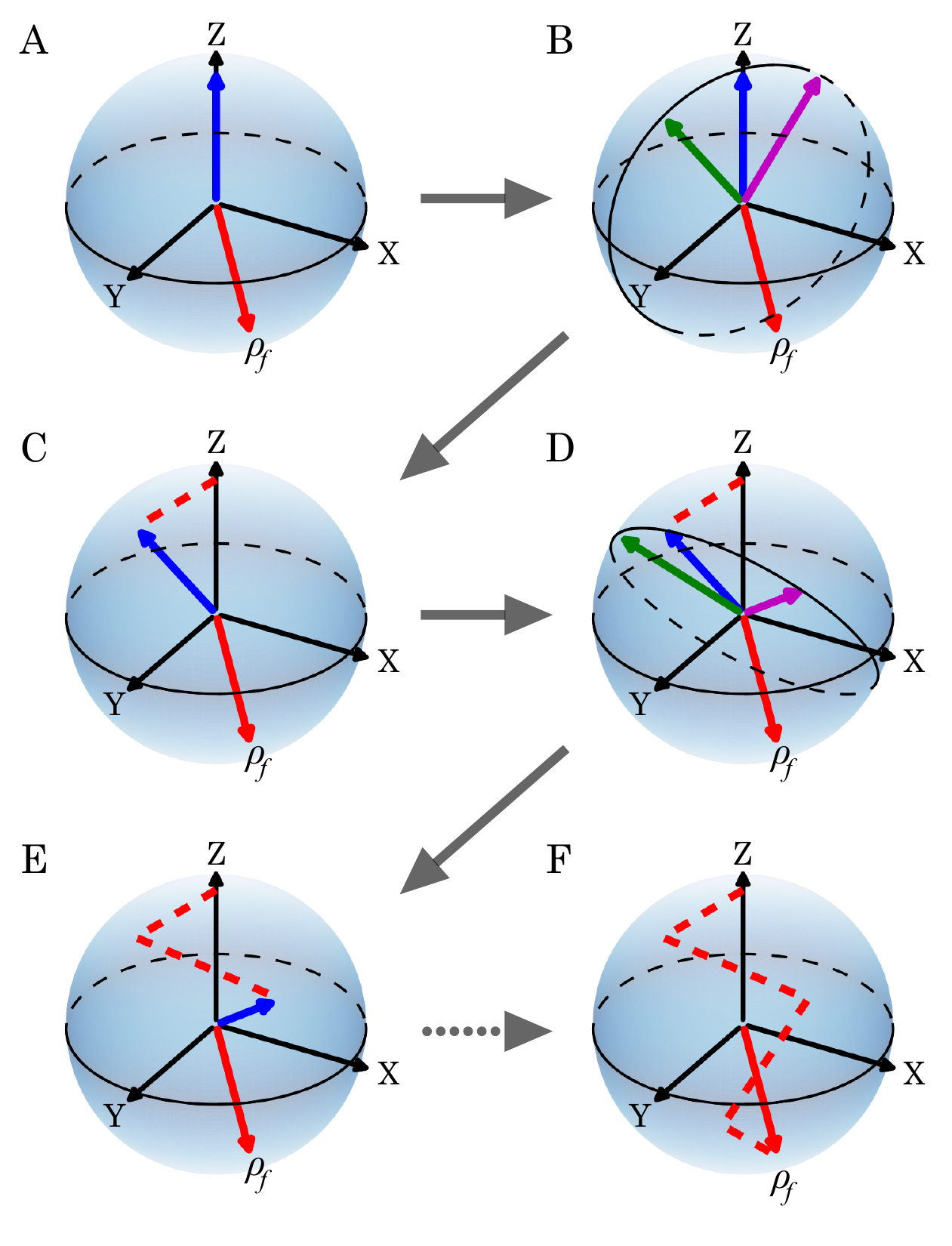}
\caption{\textbf{Self-guided quantum tomography algorithm.} (\textbf{A}) The unknown quantum state that we want to characterize is shown as a red Bloch vector and the current estimation is shown as a blue Bloch vector. (\textbf{B}) The algorithm estimates the gradient in a stochastically chosen direction by making expectation value measurements with the green and purple projectors. (\textbf{C}) The algorithm steps in the direction of the highest expectation value and the current estimate of the state is updated. (\textbf{D-F}) The gradient is estimated again and the process repeated for a set number of iterations.}
\label{figure:illustration}
\end{figure}

Adaptive quantum tomography (AQT) has demonstrated improved efficiency and precision by using state dependent tomographic measurements \cite{fischer_quantum-state_2000,huszar_adaptive_2012,hannemann_self-learning_2002,mahler_adaptive_2013,kravtsov_experimental_2013,struchalin_experimental_2015,qi_recursively_2015}. AQT relies on solving an optimization problem using previous results to select the next measurement to be performed. As a result, AQT is as computationally expensive as SQT and likewise is sensitive to statistical noise and experimental errors.

SGQT is an autonomous, robust and precise method for characterizing quantum states. Here we demonstrate the performance and robustness of SGQT in several one- and two-qubit experiments. SGQT treats tomography as a projection measurement optimization problem using an iterative stochastic gradient ascent algorithm \cite{spall_multivariate_1992}. SGQT is therefore robust against both statistical noise and experimental errors. SGQT avoids many of the pitfalls of SQT and AQT at small added cost in the number of different measurement settings required. 

SGQT iteratively learns the quantum state through maximising the expectation value of a projection measurement. The algorithm is graphically illustrated in figure \ref{figure:illustration}. The unknown quantum state $\rho_f$ is shown as a red Bloch vector and the current estimate of the state at iteration $k$ is $\ket{\phi_k}$, shown as a blue Bloch vector in figure \ref{figure:illustration}A. A direction $\Delta_k$ is stochastically chosen and the expectation values of projectors $\ket{\phi_k \pm \beta_k\Delta_k}$ are measured, shown as green and purple Bloch vectors in figure \ref{figure:illustration}B. The expectation values are measured as 

\begin{equation}
E(\rho_f,\ket{\phi_k\pm\beta_k\Delta_k}) = \bra{\phi_k\pm\beta_k\Delta_k}\rho_f\ket{\phi_k\pm\beta_k\Delta_k},
\label{equation:expectation}
\end{equation}

\noindent where $\beta_k$ controls the gradient estimation step size. The expectation value gradient in the direction $\Delta_k$ is estimated as

\begin{equation}
g_k = \frac{E(\rho_f,\ket{\phi_k+\beta_k\Delta_k})-E(\rho_f,\ket{\phi_k-\beta_k\Delta_k})}{2\beta_k}.
\label{equation:gradient}
\end{equation}

The estimate of the state is updated to $\ket{\phi_{k+1}} = \ket{\phi_k+\alpha_kg_k\Delta_k}$ in the direction of highest expectation value, where $\alpha_k$ is the step size which decreases with iteration number in order to converge. The state $\ket{\phi_{k+1}}$ is shown as a blue Bloch vector in figure \ref{figure:illustration}C and this process is repeated until termination at a set number of iterations. The final estimate of the state is the final projection $\ket{\phi_N}$ (see Supplementary Information for further details of the algorithm). The algorithm is robust against noisy gradient estimates and as such SGQT is robust against statistical noise and measurement errors.

We experimentally demonstrate SGQT using polarization encoded photonic qubits. We generate pairs of indistinguishable photons from a spontaneous parametric down conversion source \cite{burnham_observation_1970}. We prepare heralded single qubit and entangled two-qubit states using motor-controlled rotating waveplates. Projection onto any one- or two-qubit separable state is implemented using further rotating waveplates and polarizing beam splitters (see Supplementary Information for full experimental details). We calculate the expectation value in equation \ref{equation:expectation} by measuring the number of photons recorded as a proportion of the total for a fixed integration time.

\begin{figure}[t]
\centering
\includegraphics[width=1.0\linewidth]{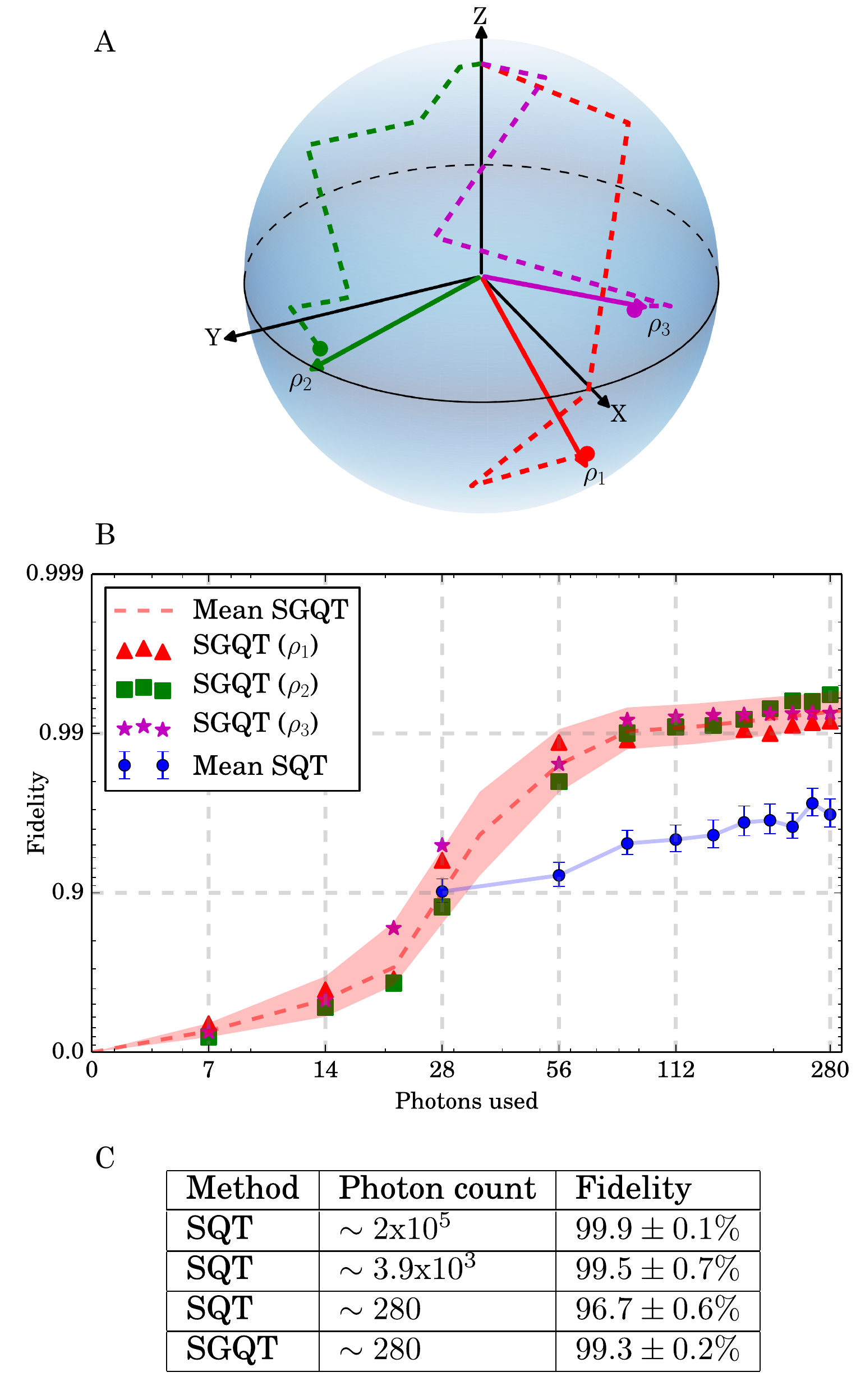}
\caption{\textbf{One-qubit SGQT performance with low count rate.} We perform SGQT in a regime where only seven photons on average are used per iteration of the algorithm. (\textbf{A}) A single SGQT route for each target state, plotted on the Bloch sphere. (\textbf{B}) The red, green and purple points show the average fidelity of SGQT for each target state. The red line shows the average fidelity across all target states. The blue points show the average fidelity of SQT across all target states. (\textbf{C}) The fidelity of SGQT with $\sim280$ photons compared to SQT with $\sim280$, $\sim3.9\mathrm{x}10^3$ and $\sim2\mathrm{x}10^5$ photons used.}
\label{figure:one_qubit_low_count}
\end{figure}

We firstly demonstrate the robustness of one-qubit SGQT against statistical noise by reducing our photon count rate such that, with the minimum integration time, we use on average seven photons. In this regime, Poissonian noise on the photon count is very high \cite{glauber_photon_1963}. We perform SGQT on three target states using the minimum integration time per iteration and repeat each run ten times. On the same target states we perform SQT with a range of integration times to control the total number of photons used, repeating each measurement ten times.

In order to benchmark our results, we obtain a high precision estimate of the target state using SQT with a long integration time to reduce Poissonian noise. The total photon count is $\sim2\mathrm{x}10^5$ and from simulation we expect a precision of $99.9\pm0.1\%$. For benchmarking we calculate the state fidelity between this high precision estimate of the state and our estimate from SGQT at each iteration. We also use this high precision estimate to benchmark SQT with varying levels of noise \cite{nielsen_quantum_????}. We emphasize that we do not expect to see optimal convergence in the fidelity to the benchmark state since our estimate is converging toward the true physical state. 

Figure \ref{figure:one_qubit_low_count}A shows the route of one SGQT run for each target state plotted on the Bloch sphere. We set the starting estimate to be $\ket{0}$, but this can be any state. Figure \ref{figure:one_qubit_low_count}B shows a log-log plot with fidelity of SGQT and SQT against the number of photons used. The red, green and purple points are the average fidelity of SGQT for each target state. The red line is the average fidelity across all target states and the band gives one standard deviation of error. The blue dots are the average fidelity of SQT across all target states. SQT on one-qubit requires $4$ measurements and therefore a minimum if $28$ photons are used. For all points, SGQT records a greater fidelity than SQT, demonstrating enhanced robustness against high levels of statistical noise. At the final iteration after $\sim280$ photons have been used, SGQT achieves a fidelity of $99.3\pm0.2\%$, whereas SQT records a fidelity of $96.7\pm0.6\%$. To reach the same fidelity, SQT requires an order of magnitude more photons. Figure \ref{figure:one_qubit_low_count}C shows a table comparing fidelity against photon number for SGQT and SQT. In a high noise regime, this result demonstrates that SGQT is far more resource efficient than SQT.

To compare robustness to one-qubit measurement errors, we perform SGQT and SQT in a regime where we have large uncertainty of the projection measurement. We engineer this level of uncertainty by applying random errors to the waveplate settings. SQT and AQT require high precision of each projection measurement setting, whereas SGQT is robust against independent measurement errors. We apply four levels of waveplate uncertainty and perform SGQT and SQT ten times each, and measure the average fidelities. These fidelities are again benchmarked against a long integration SQT without applied errors.

\begin{figure}[t]
\centering
\includegraphics[width=1.0\linewidth]{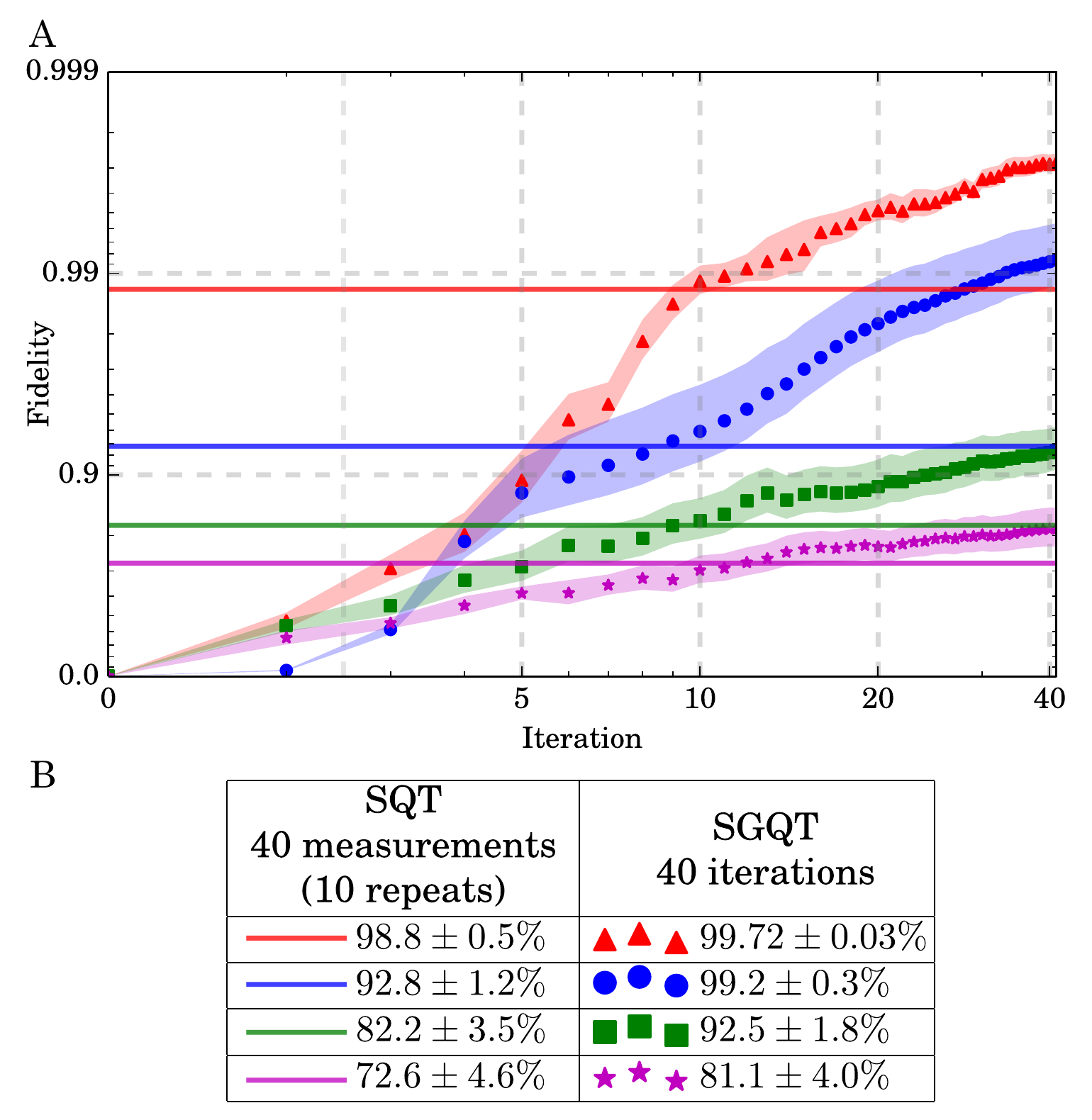}
\caption{\textbf{SGQT performance with experimental error.} (\textbf{A}) Fidelity of SGQT with varying levels of experimental error. Points are the average of ten repetitions and the band gives one standard deviation of error. SQT is performed with the same levels of experimental error, repeated ten times and the fidelities shown as solid lines. (\textbf{B}) Fidelity values for SQT after ten repetitions (40 measurements) and SGQT after 40 iterations.}
\label{figure:one_qubit_with_noise}
\end{figure}

Figure \ref{figure:one_qubit_with_noise}A presents the average fidelity of SGQT and SQT for each level of error. The results show that SGQT outperforms SQT after only $\sim10$ iterations and after $40$ iterations the infidelity (1-fidelity ($F$)) of SGQT is up to $89\%$ lower than SQT, calculated as $\tfrac{F_{SGQT}-F_{SQT}}{1-F_{SQT}}$, and will continue to decrease as numerically studied in \cite{ferrie_self-guided_2014}. These results demonstrate the robustness of SGQT to significant measurement errors. For this demonstration we reduce statistical noise by increasing the photon count rate to $\sim5\mathrm{x}10^3$ per iteration, however, in the presence of both significant statistical noise and measurement errors, SGQT will still converge with high fidelity.

We next demonstrate the performance and robustness of SGQT to characterize a two-qubit entangled state. While SGQT naturally allows for entangling measurements, in this experiments we use only local measurements which are available in our setup. We use a fidelity measure for estimating the gradient based on partial knowledge of the state. From a subset of Pauli measurements $K$, the fidelity between the two-qubit physical state $\rho_f$ and our estimate of the state $\ket{\Phi_k}$ is calculated as

\begin{equation}
\widetilde{F}(\rho_f,\ket{\Phi_k}) = \frac{1}{|M|}\sum_{i\in\{M\}}\frac{\mathrm{Tr}(\rho_fP_i^{(2)})}{\bra{\Phi_k}P_i^{(2)}\ket{\Phi_k}},
\label{equation:G}
\end{equation}

\noindent where $|M|$ is the number of Pauli measurements used per iteration and $P^{(2)} = \{\sigma_I\otimes \sigma_I, \sigma_I\otimes \sigma_X, \sigma_I\otimes \sigma_Y, \ldots, \sigma_Z\otimes \sigma_Z\}$ are the two-qubit Pauli matrices \cite{flammia_direct_2011,da_silva_practical_2011}. $\widetilde{F}(\rho_f,\cdot)$ replaces $E(\rho_f,\cdot)$ in equation \ref{equation:gradient} to estimate the gradient. In this context, existing AQT techniques would select measurements based on the solution to an optimization problem, whereas, SGQT selects a random set of measurements at each iteration and is thus much more efficient.

\begin{figure*}[t]
\centering
\includegraphics[width=1.0\linewidth]{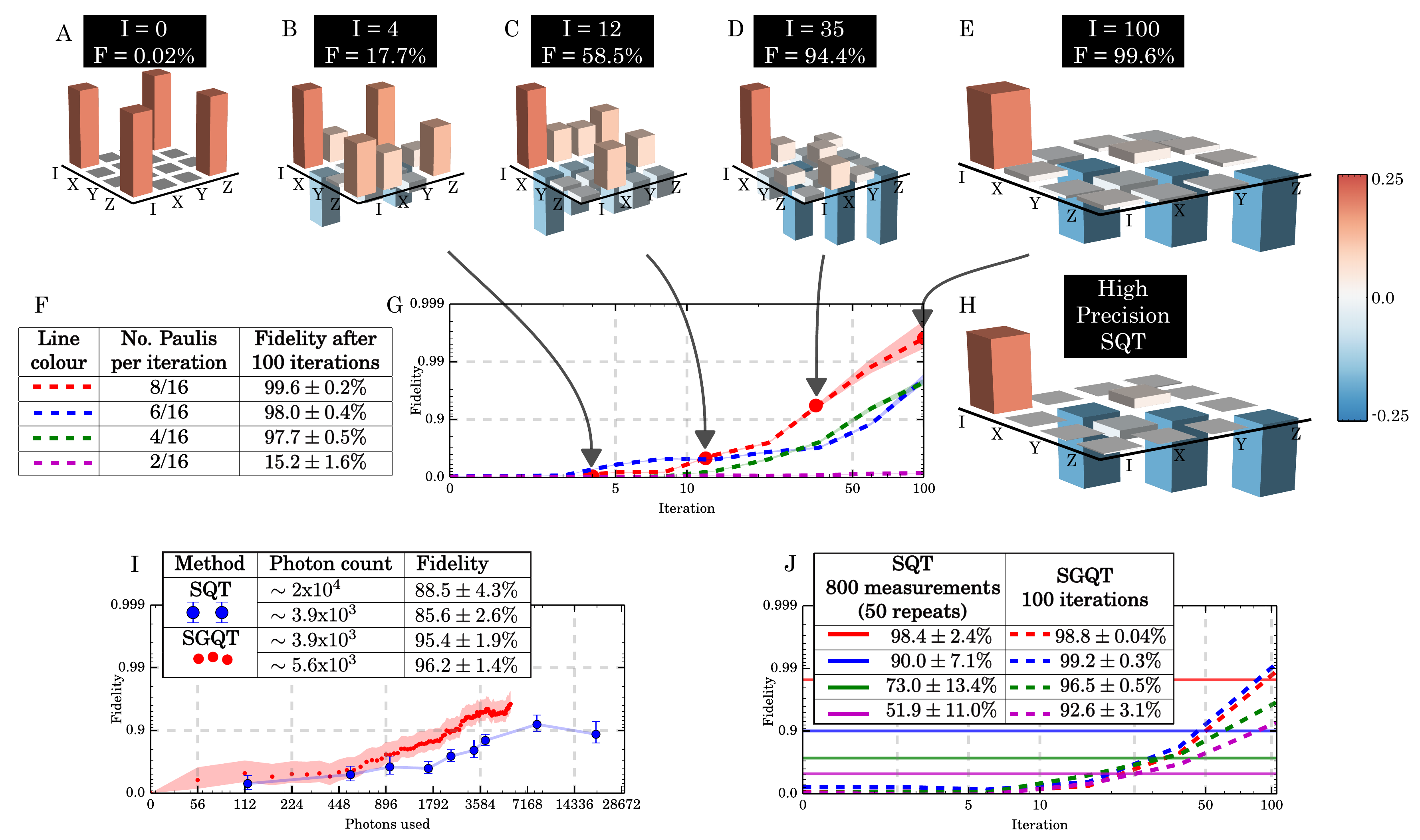}
\caption{\textbf{Two-qubit SGQT.} We perform SGQT on two qubits by taking a random subset of Pauli measurements at each iteration of the algorithm and update the state estimate based on the results. We present the state in terms of Pauli measurement expectation values, where element $i,j = \tfrac{1}{4} \bra{\Phi_k}P_i^{(1)}\otimes P_j^{(1)}\ket{\Phi_k}$ and $P^{(1)} = \{\sigma_I, \sigma_X, \sigma_Y, \sigma_Z\}$ are the one-qubit Pauli matrices. (\textbf{A-D}) The estimate of the state at different points of the algorithm run with eight measurements per iteration. The fidelity is measured against high precision SQT. (\textbf{E}) The final estimate of the state from SGQT. (\textbf{H}) The high-precision estimate of the state measured with SQT using a long integration time to reduce statistical noise. (\textbf{G}) The algorithm was run with eight, six, four and two measurements per iteration and the fidelity of the state estimate, benchmarked against long integration SQT, is plotted. (\textbf{F}) The fidelities of SGQT after 100 iterations of the algorithm. (\textbf{I}) We compare performance of SGQT and SQT in the presence of high levels of statistical noise and measure fidelity against the number of photons used. (\textbf{J}) We engineer measurement errors in waveplate rotations and measure the fidelity of SGQT and SQT with four levels of error. We repeat SQT to match the number of measurements of SGQT and average the results. Both (\textbf{I}) and (\textbf{J}) use eight measurements per iteration.}
\label{figure:two_qubit}
\end{figure*}

We experimentally demonstrate SGQT on a two-qubit maximally entangled Bell state $\ket{\Psi^-} = \tfrac{1}{\sqrt{2}}(\ket{01}-\ket{10})$. We run the algorithm taking a random subset of Pauli measurements per iteration. Figure \ref{figure:two_qubit} presents the results of the algorithm. Figures \ref{figure:two_qubit}A-E show the state estimate throughout the algorithm, presented as Pauli measurement expectation values. This run used eight Pauli measurements per iteration and figure \ref{figure:two_qubit}E shows the final estimate of the state after $100$ iterations. Figure \ref{figure:two_qubit}H shows the target state as measured with long integration SQT. The fidelity between the target state and the final SGQT estimate is $99.6\pm0.2\%$. Figure \ref{figure:two_qubit}G presents the fidelity of SGQT against iteration number for a range of total measurements per iteration. It is clear that with eight, six and four measurements per iteration (red, blue and green lines) the algorithm converges with high fidelity after $100$ iterations. With two Pauli measurements per iteration we expect the algorithm to converge, however, after a greater number of iterations. Figure \ref{figure:two_qubit}F presents the final fidelities.

We demonstrate the robustness of two-qubit SGQT against statistical noise by again reducing the photon count rate. We perform SGQT in a regime where on average seven photons per iteration are used. We also perform SQT using an equivalent total number of photons to allow direct resource comparison. Figure \ref{figure:two_qubit}I presents the fidelity of SGQT where eight measurements are used per iteration and SQT using the same total number of photons. It is clear that SGQT surpasses SQT, achieving a $68\%$ lower infidelity. SQT again requires around an order of magnitude more photons to achieve the same level of fidelity. This provides strong evidence that SGQT will outperform SQT on larger quantum systems even when only local measurements are available.

We finally investigate robustness against measurement errors on two-qubit SGQT by applying waveplate uncertainty. With the same four levels of error as the one-qubit case, we perform SGQT and SQT on the entangled Bell state. Figure \ref{figure:two_qubit}J presents the fidelity of SGQT against iteration number with experimental error applied. SQT is repeated with the same total number of measurements and the results averages. The SQT fidelities are plotter in figure \ref{figure:two_qubit}J as horizontal lines for each level of error. Using the same total number of measurements, SGQT achieves up to a $92\%$ lower infidelity than SQT.

We have demonstrated the advantages of SGQT over standard techniques for measuring quantum states in a range of one- and two-qubit experiments. When the total number of particles, in our case photons, is an expensive resource, SGQT is shown to achieve higher fidelity measurements than SQT despite the high levels of noise. SGQT is also shown to be robust against experimental errors and converges with higher fidelity than is possible with SQT measurements alone. We finally demonstrate SGQT on entangled photon pairs with high fidelity using only a random subset of Pauli measurements per iteration and is likewise robust against noise and errors.

The algorithm employed here has applications outside state tomography. This scheme can be used for state preparation, and controlling quantum devices in a similar manner could also be considered. 

While the cost of SGQT is still exponential with the system size in the number of measurements, it does not require the computationally expensive state-estimating inverse problem and maximum likelihood estimation to characterize an unknown quantum state. SGQT opens future pathways toward robust characterization of quantum systems with dimensions where standard tomographic techniques have already become impractical.

\bibliographystyle{naturemag}
\bibliography{sgqt}

\clearpage

\section*{Supplementary Material}

\subsection*{Algorithm Details}

The algorithm used in SGQT is based on simultaneous perturbation stochastic approximation (SPSA). SPSA is ideal as it requires only two noisy measurements per step in order to estimate the gradient. The main algorithm steps are described in the main text.

The algorithm parameters $\alpha_k$ and $\beta_k$ are defined as 

\begin{equation}
\alpha_k = \frac{a}{(k+1+ A)^s},\;\; \beta_k= \frac{b}{(k+1)^t},
\end{equation}

\noindent where $a,b,s,t$ and $A$ are algorithm parameters that can be asymptotically optimized offline. The asymptotically optimal values are $s=1$ and $t=\tfrac{1}{6}$, however it was often found that $s=0.602$ and $t=0.101$ performed well. In general the other parameters we kept as $a=3$, $b=0.1$ and $A=0$.

\subsection*{Experimental Setup}

Horizontally polarized photon pairs at 807.5 nm are generated via type 1 spontaneous parametric down conversion (SPDC) in a 1 mm thick BiBO crystal, pumped by an 80 mW, 403.75 nm CW diode laser. Single photons are detected in coincidence using silicon avalanche photo-diodes. To reduce noise we subtract accidental counts by measuring uncorrelated two photon events with an applied electronic delay on one channel.

\paragraph*{Heralded single polarization encoded photon source}

\begin{figure}[h]
\centering
\includegraphics[width=1.0\linewidth]{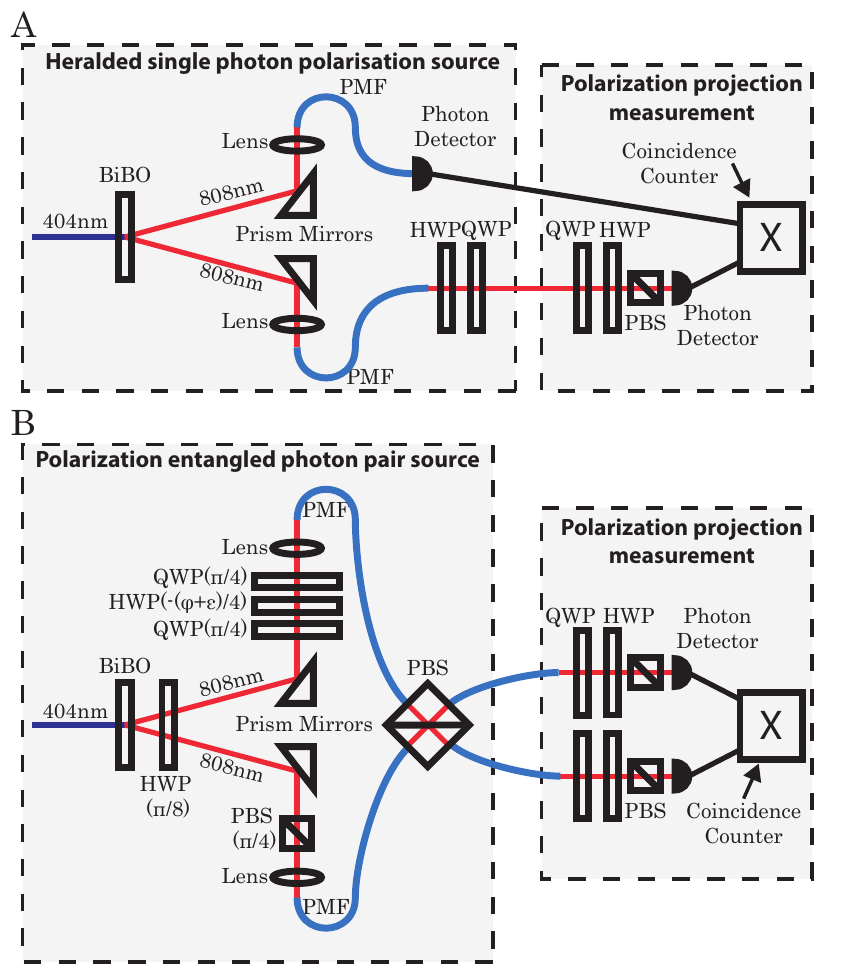}
\caption{\textbf{Experimental setup} (\textbf{A}) Photon pairs are generated by SPDC. One photon is detected directly, this heralds the existence of the other. The second photon is prepared in a polarization state by a half- and quarter-waveplate. Polarization projection measurement is performed with an HWP, QWP and polarizing beam splitter. (\textbf{B}) Polarization entangled photons are prepared by using SPDC, controlling the polarization and post selecting after both photons are incident on a PBS. Two qubit local projection measurements are performed with waveplates and polarizing beam splitters.}
\label{figure:experiment}
\end{figure}

As shown in figure \ref{figure:experiment}A, one photon is detected and triggers the presence of the other. The second photon has its polarization prepared with a half- (HWP) and quarter-waveplate (QWP). Single qubit projective measurements can be performed with a QWP, HWP and polarizing beam splitter (PBS). This enables projection onto any pure single qubit state.

\paragraph*{Polarization entangled photon pair source}

As shown in figure \ref{figure:experiment}B, both photons are rotated into a diagonal state $\tfrac{1}{\sqrt{2}}(\ket{H} + \ket{V})$ by a HWP with fast axis at 22.5$\gradi$ from vertical. One photon has a phase applied by two 45$\gradi$ QWPs either side of a HWP at $\theta\gradi$. The second photon has its diagonal state optimized with a PBS at $\sim45\gradi$. 

Both photons are collected in polarization maintaining fiber (PMF) and are incident on both faces of a fiber pigtailed PBS. When measuring in the coincidence basis, this post-selects the entangled state $\tfrac{1}{\sqrt{2}}(\ket{H_1V_2} + e^{i\phi}\ket{V_1H_2})$ where $\phi = 4(\theta + \epsilon)$ and $\epsilon$ is the intrinsic phase applied by the whole system.

PMF is highly birefringent, resulting in full decoherence of the polarization state after $\sim1$ m of fiber giving a mixed state. In order to maintain polarization superposition over several meters of fiber we use 90$\gradi$ connections to ensure both polarizations propagate through equal proportions of fast and slow axis fiber. Slight length differences between fibers and temperature variations mean the whole system applies a residual phase $\epsilon$ to the state, which can be compensated for in the source using the phase controlling HWP.

Projection onto any separable two qubit state is possible with QWPs, HWPs and PBSs.

\end{document}